\documentclass[12pt]{article}
\usepackage{epsfig,amssymb}

\newcommand{\bq}{\begin{equation}}
\newcommand{\eq}{\end{equation}}
\newcommand{\bqa}{\begin{eqnarray}}
\newcommand{\eqa}{\end{eqnarray}}
\newcommand{\ben}{\begin{enumerate}}
\newcommand{\een}{\end{enumerate}}
\newcommand{\bc}{\begin{center}}
\newcommand{\ec}{\end{center}}
\newcommand{\bqb}{\begin{eqnarray*}}
\newcommand{\eqb}{\end{eqnarray*}}

\def\gsim{\gtrsim}
\def\lsim{\lesssim}
 
\def\pr#1#2#3{ Phys. Rev. ${\bf{#1}}$ (#2) #3}
\def\prl#1#2#3{ Phys. Rev. Lett. ${\bf{#1}}$ (#2) #3}
\def\pl#1#2#3{ Phys. Lett. ${\bf{#1}}$ (#2) #3}

\def\np#1#2#3{ Nucl. Phys. ${\bf{#1}}$ (#2) #3}
\def\zp#1#2#3{ Z. f. Phys. ${\bf{#1}}$ (#2) #3}

\def\ie{{\it i.e.\/}}
\def\eg{{\it e.g.\/}}

\def\etal{{\it et.al.\/}}

\global\nulldelimiterspace = 0pt

\def\wtil#1{\widetilde{#1}}
\def\ol#1{\overline{#1}}

\def\L{ {\cal L }}
\def\O{ {\cal O }}

\def\mt{m_t}
\def\mh{m_H}

\def\lamNP{\Lambda_{NP}}

\def\Db{\ol{D}}
\def\Dbb{\ol{\ol{D}}}

\begin{document}
\pagenumbering{arabic}
\thispagestyle{empty}
\def\thefootnote{\fnsymbol{footnote}}
\setcounter{footnote}{1}
 
\begin{flushright}
PM/98-08 \\ THES-TP 98/04 \\
hep-ph/9803422 \\
March 1998 
 \end{flushright}
\vspace{2cm}
\begin{center}
{\Large\bf Testing the Higgs boson gluonic couplings at 
LHC}\footnote{Partially supported by the EC contract 
CHRX-CT94-0579.}
 \vspace{1.5cm}  \\
{\large G.J. Gounaris$^a$, J. Layssac$^b$
and F.M. Renard$^b$}
\vspace{0.5cm}  \\

$^a$Department of Theoretical Physics, University of Thessaloniki,\\
Gr-54006, Thessaloniki, Greece.\\
\vspace{0.2cm}
$^b$Physique
Math\'{e}matique et Th\'{e}orique,
UMR 5825\\
Universit\'{e} Montpellier II,
 F-34095 Montpellier Cedex 5.\\

\vspace*{1cm}
 
{\bf Abstract}
\end{center}
We study Higgs + jet production at hadron colliders in order to look
for new physics residual effects possibly described by the
$dim=6$ operators ${\O}_{GG}$ and ${\widetilde\O}_{GG}$ which
induce anomalous $Hgg$ and $Hggg$ couplings. 
Two ways for constraining these operators at LHC may be ~useful.
The first is based on the total Higgs boson production 
rate induced by gluon-gluon fusion, in which  the main cause of
limitations are due to theoretical uncertainties leading  to 
sensitivities of $|d_G|\simeq 3.\times 10^{-4}$ and 
$|\widetilde{d}_G|\simeq 1.4\times 10^{-3}$ for the 
corresponding anomalous couplings, in the  mass range 
$100~GeV \lsim \mh \lsim 2~00 GeV$. These results  imply  
sensitivity to new physics scales of  51 and  24 TeV 
respectively. The second way investigated here concerns  the
shape of the Higgs transverse momentum; for
which the theoretical uncertainties are less severe and
the limitations are mainly induced by statistics. 
A simple analysis, based on the ratio of the number of events 
at large and low $p_T$ at LHC, leads to  similar sensitivities, 
if only the $H\to \gamma \gamma $ decay mode is used. But the
sensitivities can now be improved 
by a factor 2 to 10, depending on the Higgs mass, if 
the Higgs decay modes to $WW^*$, $ZZ^*$, $WW$, $ZZ$ are also used.\\


%
%
%
\def\thefootnote{\arabic{footnote}}
\setcounter{footnote}{0}
\clearpage

\section{Introduction}

The Higgs mechanism is the cornerstone of the Standard Model (SM)
\cite{gunion}. Much experimental and theoretical effort has
already been done to discover the Higgs boson,  and 
to understand the origin of the scalar sector and constrain 
the Higgs boson mass, which is to a large extent arbitrary. 
At present, the negative experimental 
searches at LEP2 lead to the conclusion that $\mh \gsim  77$ GeV
\cite{Murray}, while  precision
measurements at Z peak give $\mh= 115^{+116}_{-66} GeV$ 
within the SM context  \cite{lepcomb}. These
values are well within the bounds obtained from purely theoretical
considerations \cite{Alt}.\par

Searches are presently going on at LEP2 and the Tevatron and 
will be pursued at LHC. If the Higgs boson is light, its
discovery should  be at hand. However, such a
discovery, will only constitute the first step in
the study of the scalar sector. Detailed studies of the properties of
the Higgs candidate should then be performed aiming at first
confirming the Higgs nature, and then  trying to get information about
the origin of the scalar sector. This ~origin may very well lie in an 
underlying fundamental dynamics generically called 
new physics (NP).\par

Assuming that the NP degrees of freedom are associated to an
energy scale $\Lambda_{NP}$ much larger than any conceivably accessible
energy range, any observable NP manifestation  
could only be in the form of residual effects modifying 
the SM interactions among usual particles
and the (yet to be discovered) Higgs boson.
At the energy range of the foreseeable colliders, the leading
such effects may arise from the set of $dim=6$  $SU(3)\times
SU(2) \times U(1)$ gauge invariant operators
\cite{Buchmuller, top-op, top-Young}, 
involving an isodoublet Higgs field and the particles
with the highest affinity to it, which inspired by SM, are taken
to be the quarks of the third family. Of course, apart from these
fields, the NP operators necessarily also involve 
gauge bosons, inevitably introduced by the gauge principle whenever
a derivative appears. The complete list of the CP conserving
such operators has been given in \cite{top-op, top-Young},
while the CP violating ones have appeared in 
\cite{top-CP, Tsirigoti-CP}.  Tests of their effects at present and
future colliders have been given for most of them in 
\cite{ top-op,top-Young,LEP1-constraints, LEP2, NLC, dijet} .\par

Among these operators, those inducing anomalous Higgs-gluon 
couplings beyond the SM ones
arise when the NP degrees of freedom are coloured. Examples of dynamical
models in which such operators are generated have been pointed out
\cite{top-op}, but very little has yet been said about their
possible signatures \cite{Mahanta}.\par

The aim of this paper is to study the Higgs-gluon couplings
induced by these operators, 
namely the CP-conserving one dubbed ${\O}_{GG}$ and 
its CP-violating analogue
$\widetilde{\O}_{GG}$, respectively associated to dimensionless
couplings $d_G$ and $\widetilde{d}_G$. These  operators 
are the gluonic analogues of the operators ${\O}_{BB}$, ${\O}_{WW}$,
$\widetilde{\O}_{BB}$ and $\widetilde{\O}_{WW}$, which are
inducing anomalous Higgs couplings to the electroweak gauge
bosons and have been studied in \cite{HZ,Vlachos, Helec,Tsirigoti-CP}. 
Since the SM couplings among Higgs bosons and
gluons, first appear only at the 1-loop level, (essentially through 
a top quark loop), and are therefore ~somewhat reduced, 
there is a favorable situation 
for the detection of any NP contribution  induced at tree level
by the possible  operators $\O_{GG}$  and $\wtil{\O}_{GG}$. 
Other types of NP effects, like \eg\@ 
the anomalous $gtt$ and $Htt$ couplings  modifying the
 $Hgg$ and $Hggg$ interactions through a top-loop contribution, 
may also appear, \cite{Diaz}. Below, we show how the presence 
of these  NP effects can be tested through studies of
Higgs boson production and decay.\par

As the $H \to gg$ branching ratio
only represents a small fraction (6-7\%  for 
$\mh \simeq 100-150$ GeV and
much less for higher masses), accurate measurements of the $Hgg$
couplings through  decay width $\Gamma (H\to gg)$ 
will need a copious Higgs boson
production and very powerful methods for distinguishing $gg$ from
light quark final states. 
At $e^+e^-$ colliders, only a few $H \to gg$ events are
expected through $HZ$  production; \ie\@
a few tens at LEP2 and a few hundreds at a higher energy linear
collider (LC), depending on the
achieved luminosity. An analysis based on a luminosity
of $50~fb^{-1}$ at a $500~GeV$ LC, led to an uncertainty of $\pm39\%$
for the sum $c\bar c+gg$ \cite{Hildreth}. 
The same situation probably also  
arises for $H$ production in a $\gamma \gamma $ Collider.
At the upgraded Tevatron, a hundred of Higgs particles 
should be produced (mainly through the
process $p\bar p\to WH+X$) for $\mh \simeq 100-150$;
but its observation seems possible
only in the $b\bar b$ mode \cite{GunionH}.\par 

The situation should be quite different at LHC where 
a copious Higgs production  of about
$10^5-10^6$ events per year for $ \mh \lsim 200 GeV $,  
should be expected  \cite{H4, Dawson, ellis, spira1, spira2}.
At the LHC  energies, the largest cross sections arise from  
subprocesses with 
$gg$, $WW/ZZ$ and $q\bar q$ initial states, as well as from 
bremsstrahlung off a top quark. Among these, the process $ gg \to H$
largely dominates. So
this should be the best place to look for anomalous $Hgg$, 
$Hggg$ couplings. The standard prediction for this process 
has been computed and found very sensitive to higher order  
QCD corrections due to soft gluon effects, for the resummation
of which there exist large theoretical uncertainties  
\cite{spira, kramer}. This means 
that an accurate measurement of $Hgg$ couplings 
through the total production rate is only hindered by theoretical
uncertainties, while statistics are enormous. 
Assuming a conservative
theoretical error for the QCD  corrections, we estimate
the discovery limits for the ${\O}_{GG}$ and
$\widetilde{\O}_{GG}$ effects. These limits are interesting, 
since they correspond to NP scales
lying in the tens of TeV range.\par 

A theoretically cleaner  (and in any case complementary) 
study of the operators $\O_{GG}$ and
$\wtil{\O}_{GG}$ could be
achieved by looking at the Higgs ~transverse momentum distribution
in the {\it Higgs + jet} production process. To our knowledge, this
has never been discussed before for the search of NP effects. 
To achieve this, one first needs the 1-loop SM amplitudes for 
$gg \to Hg$, $gq\to Hq$
and $q\bar q\to Hg$,  that will interfere with the NP ones.
These SM contributions have been  computed in 
\cite{Hjet, RKEllis}. We have recomputed them and checked
numerically the agreement with
the results obtained previously. Adding then the NP contributions
due to the operators ${\O}_{GG}$ and $\widetilde{\O}_{GG}$, we have
examined how the SM predictions for the various observables,
(like \eg\@ the distributions of the Higgs rapidity and
transverse momentum, and the Higgs+jet
invariant mass and angular distributions)
are influenced by NP. \par

The most striking effect is, as expected, in
the $p_T$ dependence. The SM prediction (as well as the one of 
any other NP model
affecting only $gtt$ and $Htt$ couplings) should drop down as soon as 
$p_T \gsim m_t$; whereas the NP contribution 
due to ${\O}_{GG}$ and ${\widetilde\O}_{GG}$ operators
(associated
to a heavy $\Lambda_{NP}$) stays flat, leading to a clear
signal for anomalous couplings. We then derive the observability
limits on $d_G$ and $\widetilde{d}_G$ by considering 
the ratio of Higgs+jet production rates, at the high and low
$p_T$ regions. \par

The content of the paper is the following. In Section 2 we present the
${\O}_{GG}$ and ${\widetilde\O}_{GG}$ operators and  derive the
unitarity constraints relating the NP couplings to
the scale $\Lambda_{NP}$, where the new degrees of freedom
start being excited. We then present the formulae for the 
NP and SM contributions to the $H\to gg$ decay, and to the 
Higgs boson and $H+~jet$ production in $pp$ collisions. 
The sensitivity limits
to the NP couplings are derived in Section 3, first on the basis of the
expected accuracy in the measurement of the $\Gamma (H\to gg)$  width; 
and then on the basis of the Higgs production rate and the
various $H+~jet$ distributions at LHC. We find that for $100
\lsim \mh \lsim 200 GeV$, the study of the total Higgs production
rate, and the study of the $p_T$-distribution in the case that 
only  $H \to \gamma \gamma$ is used for $H$-detection, give comparable
results. On the other hand the  $p_T$-distribution technique starts 
becoming superior whenever the $H \to WW^*, ~ ZZ^*,~ WW, ~ZZ$ 
decay modes are  also used to increase its statistics, which 
may even lead to an order of magnitude improvement
as $\mh$ approaches the 200 GeV region.
Section 4 summarizes the results and their 
implications for the NP search. 
Technical details on invariant amplitudes, loop computations and parton
kinematics are collected in an appendix.

\section{Formalism.}
\subsection{The NP operators ${\O}_{GG} $ and $\widetilde{\O}_{GG}$.}
We consider the NP effects arising from the effective
Lagrangian
\bq
 \L_{NP} ~= ~\frac{d_G}{v^2} {\O}_{GG} 
+ \frac{\widetilde{d}_G}{v^2}\widetilde{\O}_{GG}\ \ ,
\label{NPlagrangian}
\eq
where  $d_G$ and $\widetilde{d}_G $ are NP dimensionless couplings,
and the NP operators are
\bqa
{\O}_{GG} &
= & \left ( \Phi^\dagger \Phi -\frac{v^2}{2} \right )
\overrightarrow G_{\mu\nu} \cdot \overrightarrow G^{\mu\nu}
 \ \ \  , \ \  \label{OGG} \\   
{\widetilde\O}_{GG}
& = & \left (\Phi^\dagger \Phi \right )\overrightarrow
{\widetilde G}_{\mu\nu} \cdot \overrightarrow G^{\mu\nu}
\ \ \   , \ \  \label{OGGtilde}   
\eqa
\noindent
with $\wtil{G}^i_{\mu \nu} \equiv
1/2\epsilon_{\mu\nu\lambda\sigma}G^{i\lambda \sigma}$ and
\begin{equation}
\Phi=\left( 
\begin{array}{c}
i\chi^+ \\ 
{\frac{1}{\sqrt2}}(v+H-i\chi^3)
\end{array}
\right) \ \ \ \ . \ \ 
\end{equation}\par

Unitarity allows to establish for each operator an unambiguous
relation between the  NP coupling
constants appearing in (\ref{NPlagrangian}),
and the corresponding energy scale $\lamNP$, 
at which unitarity is saturated.  This scale 
supplies a practical definition of the
scale  where the new physics  generates the
corresponding operator \cite{top-op, unitarity, Vlachos}.
\par  

In the case of the ${\O}_{GG} $ and $\widetilde{\O}_{GG}$
operators, the strongest  unitarity constraint arises from
the $J=0$ partial wave transition amplitude 
affecting the colour singlet channels  
$|gg\pm\pm \rangle$ and $|HH\rangle$.  
Using (\ref{NPlagrangian}), we thus get
for ${\O}_{GG} $ the unitarity relation 
\bq
d_G ~=~{4\pi\over1+c\,{\pi v^2\over\lamNP^2}} \left (
{v^2\over\lamNP^2} \right )  \ \  \ 
\label{DGunitarity}
\eq 
\noindent
with $c=31$ for $ d_G>0$, and $c=-1$ for $ d_G<0$. The unitarity
relation for  $\widetilde{\O}_{GG}$ is
\bq
|\widetilde{d}_G| ~=~{4\pi\over1+15\,{\pi v^2\over\lamNP^2}} 
\left ({v^2\over\lamNP^2} \right )  \ \ . \label{DGTunitarity}
\eq\par

\subsection{The $H\to gg$ decay width}

In this section we give the modification to the Higgs gluonic decay 
width induced by tree level effects of the operators
${\O}_{GG}$ and $\widetilde{\O}_{GG}$. \par

The $Hgg$ gauge-invariant amplitude\footnote{The phase
of the amplitude is defined to be that of the $S$-matrix
element.} $R(Hgg)$ is given by
\bq
R(Hgg)~= ~i f(k^2)\delta_{ab}
[(k.k')(\epsilon.\epsilon')-(\epsilon.k')(\epsilon'.k)]
\ \ , \ \label{RHgg} 
\eq
where the gluon momenta and polarization vectors are denoted as
$(k,~\epsilon)$ and $(k',~\epsilon')$ respectively\footnote{All 
momenta are taken as incoming.}. In (\ref{RHgg}), the
momentum $k$ is allowed to be off-shell, while the other gluon with 
momentum $k'$ and polarization $\epsilon'$ is always on-shell.
Under such conditions, there is only one gauge invariant form for
$Hgg$, which is shown in (\ref{RHgg}). 
The indices $(a,b)$ specify the colours of the two
gluons. In SM, the dominant contribution arises from the top 
triangle in Fig.1a  and is given by
\bq
f(k^2)={ \alpha_s  m^2_t \over \pi v }\bar{f}(k^2) \ \ 
\ , \ \label{fk2}
\eq
in terms of $\bar{f}(k^2)$ presented in Appendix A1. 
Here we only note that $f(0)$ agrees with the result quoted
in \cite{ellis, spira1, spira, Hjet, RKEllis}. In the heavy
quark limit 
$m_t \gg  \mh$, this leads to the well known result
\cite{Georgi, RKEllis}
\bq 
\bar{f}(0) \to -\, {1\over3m^2_t} ~~.
 \eq  \par

In the presence of the NP contribution given in (\ref{NPlagrangian}),
the Higgs  decay width into 2 real gluons is given by
\bq 
\Gamma(H\to gg)= {\mh^3 \over8\pi}\left 
\{|f(0)-{4d_G\over v}|^2+{16\over v^2}
|\widetilde{d}_G|^2 \right \} \label{Hgg} \ \ .
\eq
As expected, only the CP-conserving NP contribution interferes 
with the SM one.

\subsection{Higgs and Higgs+jet production at pp colliders}
\subsubsection{Single Higgs production}

At lowest QCD order, including the effect of the
NP operators ${\O}_{GG}$ and $\widetilde{\O}_{GG}$, 
the total rate for $pp\to H + X$ 
due to the subprocess $gg\to H$
is given by
\bq
\sigma^0(pp \to H+X)= \hat{\sigma}^0\tau_H L_{gg} \ \ ,
\label{sig0}
\eq
where 
\bq
 L_{gg}=\int^1_{\tau_H} {dx\over x}
g(x)g\left ({\tau_H\over x}\right ) \ \ ,
\label{sig01}
\eq
\bq
\hat{\sigma}^0~ =~ \frac{\pi^2}{8\mh^3} \Gamma(H\to gg)
~ =~ {\pi \over64}\left \{|f(0)-{4d_G\over v}|^2+{16\over v^2}
|\widetilde{d_G}|^2\right \} \ \  , \ \ \
\label{sig02} 
\eq
\noindent
$ \tau_H=\mh^2/ s$ and 
$g(x)$ is the gluon distribution function inside the proton.\par

QCD corrections correspond to including loop corrections to
$gg \to H$, as well as corrections due to associated production
of massless partons together with the Higgs, in the processes 
$gg\to Hg$, $q\bar q\to Hg$, $gq\to Hq$, $g\bar q\to H\bar q$
\cite{H4, Dawson, H8, spira1, kramer, spira, H7}. The  result in SM 
is  
\bq
\sigma(pp\to H+X)=\hat{\sigma}^0 
\left [1+C \frac{\alpha_s}{\pi} \right ]
\tau_H L_{gg}
+\Delta\sigma_{gg}+\Delta\sigma_{gq}+\Delta\sigma_{g\bar q}
+\Delta\sigma_{q\bar q}
\label{sig03}
\eq
where the various terms correspond to $gg$, $gq(\bar q)$ and 
$q \bar q$ initial state. Depending on the value of the Higgs
mass, these QCD corrections increase the Higgs production rate
by 60\% to 90\% \cite{H4, H8, spira1, spira, kramer, H7}.\par

\subsubsection{H+jet production at large $p_T$.}

We now turn to Higgs + jet production (Fig.2), which at 
a hadron collider, takes place  through
the subprocesses (Fig.3) $gg\to Hg$,  
$gq\to Hq$, $g\bar q\to H\bar q$,  $q\bar q\to Hg$
\cite{Hjet, RKEllis}.\par

We have repeated the computation of \cite{RKEllis}  
of the triangle and box
contributions to the various types of subprocesses participating
in the $H + ~jet$ production (compare Fig.3), and we have added 
the tree level NP contributions due
to ${\O}_{GG} $ and $\widetilde{\O}_{GG}$. The ${\O}_{GG}$
contribution 
interferes with SM, whereas the $\widetilde{\O}_{GG}$ one does not and
adds quadratically in the cross section.\par

Details on the amplitudes are given in Appendix A2; while
those on the kinematics of the 
two-body inclusive parton model distributions, 
are presented in Appendix A3. We collect there the
expressions for the Higgs transverse momentum
distribution $d\sigma/dp_T$, the $H$ rapidity distribution $d\sigma/dy_H$,
the (H+jet) invariant mass distribution
$d\sigma/dM$, and the angular distribution
$d\sigma/d\chi$. They are obtained by convoluting the elementary
differential cross section $d\hat{\sigma}/d\hat{t}$ for the
various subprocesses, with gluon and quark distributions.\par

For $gg \to Hg$ the elementary differential cross section 
is 
\bqa
{d\hat{\sigma}(gg\to Hg)\over d\hat{t}} &= & 
\frac{3 \hat u \hat t}{128^2 \pi \hat s}
\Bigg \{ (\hat u \hat t)^2
\left (|A_1|^2+ {\hat s^2 \over 2}|A_4|^2 \right )
+4(|A_2|^2 \hat u^2+|A_3|^2 \hat t^2)
\nonumber\\
&&-4 \hat{u} \hat{t} 
 \Re(A_2A_3^*)+
2\hat{u}\hat{t}\Re((\hat t A_3-\hat u A_2)(A_1^*+\hat s
A_4^*))+\hat{s}(\hat t \hat u)^2\Re(A_1A_4^*) \Bigg \}\nonumber\\
&&+~{3\alpha_s |\widetilde{d}_G|^2\over8v^2}
\left [{2p^2_T \over \hat s^2}
-{4\over \hat s}(1-{M^2_H\over\hat s})^2
+{1\over p^2_T}\left ((1-{M^2_H\over\hat s})^4+1+({\mh^2\over
\hat s})^4\right )
\right ] \ . \label{sigmat-ggHg}
\eqa
In this expression the functions $A_i$ contain both SM and ${\O}_{GG}$
contributions as given by (\ref{ap-A1},
\ref{ap-A4},\ref{ap-A2},\ref{ap-A3}) in the
Appendix A2. The quantity $p_T$ denotes the 
transverse Higgs momentum; 
compare Appendix A3.
As expected, the CP-violating ${\widetilde\O}_{GG}$
contributions does not interfere with SM and appear separately in
quadratic form. This $\wtil{d}_G$ quadratic term 
is the same as the quadratic term  of the
$\O_{GG}$ contribution (coming from the functions $A_i$) by 
just replacing $d_G \to \widetilde{d}_G$.\par

While for $q \bar q \to Hg$ and $g q(\bar q) \to H q (\bar q)$
we get
\bq
{d\hat{\sigma}(q\bar q\to Hg)\over d\hat{t}}~ = ~ 
{\alpha_s\over36\hat{s}^3}(\hat{t}^2+\hat{u}^2)
\Big \{|f(\hat{s}) -{4d_G\over v}|^2
+{16\over v^2}|\widetilde{d}_G|^2 \Big \} \ \ ,
\label{sigmat-qqbar}
\eq
\bq
{d\hat{\sigma}(gq\to Hq)\over d\hat{t}} ~ = ~
{d\hat{\sigma}(g\bar q\to H\bar q)\over d\hat{t}} ~ = ~
{\alpha_s\over96\hat{s}^2|\hat{t}|}(\hat{s}^2+\hat{u}^2)
\Big \{|f(\hat{t})-{4d_G\over v}|^2
+{16\over v^2}|\widetilde{d}_G|^2 \Big \} \ \ .
\eq
The corresponding expressions for $\bar q q \to H
g$, $qg\to Hq$, $\bar qg\to H \bar q$ are obtained through the
replacement $ \hat t \Longleftrightarrow \hat u$. 
The function $f(k^2)$ representing the Standard Model Hgg triangle
loop contribution, has been defined in (\ref{fk2}) and in Appendix A1.

\section{Sensitivity to NP couplings}

\subsection{From $\Gamma(H\to gg)$ measured at Linear Collider.}

We  give an estimate of the sensitivity to the NP couplings
$d_G$ and $\widetilde{d}_G$ for a light Higgs boson 
($\mh \simeq 100-200 GeV $), assuming a global uncertainty on
$\Gamma(H\to gg)$ of about 40\%. This would cover the theoretical
uncertainties on $\alpha_s$ and higher QCD effects, as well as
the experimental errors in the measurement of this decay width.
It seems that even with a high energy,
high luminosity $e^+e^-$ Linear Collider (LC), 
this is the best one can expect \cite{Hildreth}. With
this assumption one obtains the sensitivity limits
\bq
|d_G| ={v\over20}|f(0)| \ \ \ \ , \ \ \ \ \ 
|\widetilde{d}_G|^2 ={v^2\over40}|f(0)|^2
\eq
which, using also the unitarity constraints (\ref{DGunitarity},
\ref{DGTunitarity}), means \footnote{Note that our notation for
the NP scale is different from the one in  \cite{Mahanta}.}
\bq
|d_G| = 6.\times10^{-4}~~~~~  (\lamNP=36~TeV) \ \ \ , \ \ \ 
|\widetilde{d}_G| = 2.\times10^{-3} ~~~~~~ (\lamNP=20~TeV) \ . 
\eq

These would be already quite remarkable values. They arise because 
one compares tree
level NP effects with 1-loop SM contributions.\\

\noindent

\subsection{From the Higgs production rate at LHC.}

At LHC, Higgs production should be dominated by the $gg\to H$
process. We therefore consider the production rate 
in $pp\to H+X$ given by 
(\ref{sig0} - \ref{sig03}).
Theoretically, the  sensitivity to NP 
should be similar to the one expected
from $\Gamma(H\to gg)$, as it is this same quantity which controls
the production rate. 
Experimentally,  this Higgs production process is very
interesting, due to the large number of events expected in the various
Higgs decay modes. For a light Higgs ($\mh \simeq 100-150$ GeV)
the $\gamma\gamma$ channel is
experimentally favored (since the statistics is enormous anyway),
and the expected experimental accuracy is of a few percent 
\cite{GunionH}.\par 

In this analysis we assume that the NP effect 
on $Br(H\to\gamma\gamma)$ is negligible. This stems from the
observation that it could only come from two possible sources.
The first one is a direct NP effect on the $H\gamma\gamma$
coupling arising from other  NP operators 
(not involving gluon fields), which can be constrained through
different processes though, as discussed 
in \cite{HZ,Vlachos, Helec,Tsirigoti-CP}.
The second possible source comes from a  modification of
the $\Gamma(H\to gg)$ width induced by $\O_{GG}$, $\wtil{\O}_{GG}$,
which in turn changes  accordingly the total $H$-width
and  $Br(H\to\gamma\gamma)$. However, as the
$Br(H\to gg)$ is only a few percent, even a  20\% modification 
of $\Gamma(H\to gg)$, induces only a $\sim 1\%$ variation to 
$Br(H\to\gamma\gamma)$, which is negligible  compared
to the experimental uncertainty.
In addition, to the already mentioned theoretical uncertainty in the 
QCD corrections, one should add the uncertainty in the parton
(mainly the gluon) distribution functions. In
\cite{spira, kunszt, MRSTpartons},
these are estimated to be of the order of $\pm 10~ \%$ to
$\pm 20~ \%$. So they turn out to be the dominant source of
uncertainty in this analysis.
Assuming a global uncertainty of $\pm 20~ \%$ will
give the sensitivities: 
\bq
|d_G| \simeq  3.\times 10^{-4}~~~~~  (51~TeV) \ \ \ \ , \ \ \ \ 
|\widetilde{d}_G| \simeq  1.4\times 10^{-3} ~~~~~~ (24~TeV) 
\label{d-rate}
\eq
\noindent
Numerically, these numbers are only slightly better than the
ones given in the preceding
Section on the basis of $\Gamma(H\to gg)$ at LC. But since 
LHC should anyway run before LC, we must 
emphasize  the importance of these results,
which give indeed very interesting constraints on the NP
effects.  \par

In concluding this subsection we note that the main difficulty 
here comes from theoretical uncertainties, which prohibit us to 
make a better benefit of 
the huge statistics
brought in by LHC. This is why in the next Section, we turn to a
theoretically cleaner NP signal, free of the normalization 
uncertainties of the total cross section.\par

\subsection{From the shape of the H+jet
distributions at LHC.}

We now look at  the  NP signal based on  the relative (but drastic)
differences in the shape of the  $H$ production
at large transverse momentum.\par

At $p_T \gsim m_t$, the SM distribution
(based on the triangle $Hgg$ and the box $Hggg$ contributions due to the top
quark loop in Fig.1) starts ~being sensitive to the non-locality
of the $Hgg$ vertex and falls off; whereas the NP contribution 
(associated to a large scale $\Lambda_{NP}$) is still local and
remains flat. In fact, one can check that the pure NP 
contribution behaves in the same way as the contribution  called
$SM_{eff}$ in Figs.4-7 below, which is derived 
by taking {\it  the large $m_t$ limit} in SM  
 and identifying $|d_G| = {\alpha_s\over12\pi}$ or 
$|\widetilde{d}_G| = {\alpha_s\over12\pi}$, 
\cite{Dawson, Dawson1}.\par

The results for various  $(H + jet)$ distributions
are shown in Figs.4-7. In all our illustrations $\mh=100~GeV$ is
used, but we have checked that there is very little change 
when this mass increases up  to  200 GeV.
In Figs.4-7, we compare the complete 1-loop SM
prediction (labeled  $SM$), with the  large $m_t$
approximation to SM (labeled  $SM_{eff}$), and  
the effects due to the presence of  
the ${\O}_{GG} $ operator  with
$d_G=+ 10^{-3}$ (labeled  $+d_G$) or $d_G=- 10^{-3}$
(labeled $-d_G$), as well as the effect of the   $\widetilde{\O}_{GG}$
contribution for $|\widetilde{d}_G|=10^{-3}$ (labeled
$\wtil{d}_G$). The results include all possible subprocesses 
due to $gg$, $gq ~g\bar q$ and $q \bar q$ initial states;
compare  (\ref{Sgg}-\ref{Sqqbar}). 
As shown in Fig.6, the subprocess $gg\to Hg$ dominates the $p_T$
distribution for $x_T \equiv 2 E_{TH}/\sqrt s \lsim 0.1$; 
while for $x_T\gsim 0.1$ it is
$gq, g\bar q$ that dominate; the $q\bar q$ contribution being
always smaller.  \par

The main features concerning the 
NP observability are discussed below:\\ 

{\bf Shape of $y_H$ and $\chi$ distributions.}\\
 The shape of the $y_H$ and $\chi$ (compare (\ref{chi}) 
distributions in Figs.4,5) do not seem
to be notably different in the SM and in the NP cases. They only differ
in absolute magnitude, roughly in the same way as the total production
rate. So, from these measurements, we cannot expect an improvement 
in the determination of the NP
couplings as compared to the one obtained from the total production
rate.\\

{\bf Shape of $x_T$ and $M$ distributions.}\\
On the contrary, the transverse energy and the $H+jet$ invariant
mass distributions
(Fig.6,7) are very sensitive to NP. At large $x_T$ or $M$, the NP
contributions differ from SM by a flattening of the distributions.
As already stated, such behaviour is due to the locality of the
NP interaction implied by its high $\lamNP$ scale. For the same
reason, this is 
also true for the $SM_{eff}$ approximation 
describing the large $m_t$ limit of the SM,
which should thus become inadequate  for $p_T \gsim \mt $.
The size of the NP effect, is
fixed by the value of the $d_G$ and $\widetilde{d}_G$ couplings,
which once they are determined also fix the accessible values 
of $\lamNP$ through the unitarity relations (\ref{DGunitarity},
\ref{DGTunitarity}).\par

We also remark that the quadratic $\widetilde{\O}_{GG}$
contribution always increases the rate, whereas
the linear ${\O}_{GG} $ one produces a constructive or destructive
interference with SM, depending on the sign of $d_G$. The observation of
a destructive effect would be a clear indication for ${\O}_{GG} $.\\

{\bf Ratios and sensitivity to NP}\\
As already stated, this sensitivity is based on the change in the shape 
of the $p_T$ distribution induced by NP.  
In order to quantify it, we
consider the ratio 
\bq
R=N_{High}/N_{Low} \ \ , \
\label{Ratio}
\eq
where $N_{High}$ and
$N_{Low}$ are  the number of $(H+~jet)$ events in the high
and the low transverse energy domains 
$[x_i, ~x_{Tmax}]$ and $[x_{Tmin}, ~x_i]$, respectively; compare
Fig.6. The lowest and highest points 
$x_{Tmin}$, $x_{Tmax}$  are chosen as 
$x_{Tmin}=0.0257$ and $x_{Tmax}=0.25$; but the results are
independent of their precise values. Note in particular that 
$x_{Tmin}=0.0257$ corresponds to $p_T\simeq 150~GeV$, for which 
the use of the leading QCD expression  should be
adequate. The intermediate value  $x_i$ is chosen 
in order to maximize the sensitivity of $R$ to NP effects. It
naturally turns out that $x_i$ lies close to the value where the 
$SM_{eff}$ prediction crosses the exact SM one. We  choose
therefore $x_i= 0.05$. \par

As seen in Fig.6,  the 
number of events in the  low $x_T$ domain is determined solely by the 
SM contribution; while the NP effects mainly influence the
events in the high   $x_{T}$ domain.  
This is  even more pronounced for  $x_{T}>0.25$, but
there are very few events there anyway. 
The NP observability limit is then defined  by demanding
that the NP effect on the ratio $R$ is larger than  the
$1 \sigma$ statistical fluctuations of it. \par

At LHC, with an integrated luminosity $L=100 ~fb^{-1}$ per year, 
and 3 years of running for  the two ~experiments ATLAS and CMS, we find
$N^{SM}_{High}=22320$ , $N^{SM}_{Low}=417000$ at the two 
$x_T$ domains. This should 
a priori lead to an excellent sensitivity to NP. However these
numbers are reduced by the branching ratios of the Higgs boson to
the observable channels and by detection efficiencies. For the mass
range $100~ GeV \lsim \mh \lsim 200~ GeV$, the relevant 
subprocesses are $gg\to H\to\gamma\gamma$, $gg\to H\to WW^*$ 
and $gg\to H\to ZZ^*$  \cite{GunionH}. If we only use 
the $H \to \gamma\gamma$ mode, whose branching ratio 
is very low (about $2\times 10^{-3}$ for $\mh \lsim 150~GeV$ and
negligible for higher masses),  we obtain 
$N^{SM}_{High}=22$ , $N^{SM}_{Low}=420$. With such numbers the
statistical uncertainty on the ratio R is about 25\%, which 
leads to sensitivity limits for the NP couplings similar to 
those given in (\ref{d-rate}).\par

This result can be substantially improved for most of the mass range
$100~GeV \lsim \mh \lsim 200~GeV$, 
by using also the Higgs decay modes to
$WW^*$, $ZZ^*$ which strongly increase with $\mh$. Thus   
the $WW^*$ ($ZZ^*$) modes reach \eg\@ at $\mh \sim 150~GeV$  
the level of 50\% (5\%) respectively. Above the corresponding
thresholds, the $WW$ and $ZZ$ modes will of course ~dominate 
and greatly  increase the statistics. In \cite{GunionH} an
analysis of the two-weak boson final states is performed, which
shows that  their statistical error is already smaller than the one for 
the $\gamma\gamma$ case, in most of the above mass range. 
Considerable work should of course still be done
on the detection of the various Higgs decay modes, before we
fully identify all potentialities \cite{GunionH,
spira}. Using though the existing results  we are
led to the conclusion that  an improvement of the sensitivities 
given in (\ref{d-rate}) by a factor of 2 to 10, depending on
$\mh$, should be possible. This is
particularly  true for $\mh \gsim 120 GeV $. \par

We should also note that our present study of the $p_T$ shape 
through  the ratio $R$, is intended only as  a preliminary orientation. 
In the actual search, a global study of the shape of the 
$p_T$ spectrum should 
be done in a more precise way during
the event selection, taking into account the $p_T$ dependence of the
background and all characteristics of the detectors
\footnote{A recent study\cite{Abdullin} of the process 
$pp\to\gamma\gamma+jet$
shows that the background has indeed different distributions.}. 
This is beyond the
scope of this paper and our competence. We nevertheless believe that
our simple study has shown that it is reasonable to expect an
appreciable improvement in the sensitivity to the NP contribution.\par

Finally we note that we obtain similar results by applying 
the same procedure to the invariant mass distribution. 
The intermediate value which optimizes the sensitivity is now found
as $M_i \simeq 0.9 ~TeV$. The results for
$d_G$ and $\widetilde{d}_G$ sensitivities turn out be similar to the
ones obtained from  the $x_T$ distribution.\par

\section{Conclusion}

In this paper we have studied the $dim=6$ 
$SU(3)\times SU(2)\times U(1)$ gauge
invariant 
operators ${\O}_{GG}$ and $\widetilde{\O}_{GG}$, which induce
anomalous Higgs boson gluon interactions. Such effects should appear
when the NP degrees of freedom carry colour and simultaneously couple
to the Higgs sector. For studying the anomalous $Hgg$ and $Hggg$
couplings, we have looked at the tests which can be realized from Higgs
production and decay. \par

At $e^+e^-$ colliders, the measurement of the $Hgg$ coupling 
has to rely on the difficult measurement of
the $H\to gg$  width. At LEP2 only a few events could be
observed. At a 500 GeV LC, for $\mh \simeq 100-150$ GeV,
with a few hundreds of events accessible, one can only expect to reach the
sensitivity limit
\bq
|d_G| \simeq 6.\times10^{-4}~~~  (\lamNP \simeq 36~TeV) \ \ \  , \ \ \ 
|\widetilde{d}_G| \simeq  2.\times10^{-3} ~~~ (\lamNP \simeq 20~TeV) 
\ \ .
\label{d-LC}
\eq

At LHC, the $ggH$
coupling controls the $H$ production rate. Statistics is huge. 
The main problem in this case is due to 
the large uncertainties (about 20\%)
in the QCD corrections to the total cross section. 
This limits the sensitivity to NP
to the values:
\bq
|d_G| \simeq 3.\times10^{-4}~~~  (\lamNP \simeq 36~TeV) \ \ \  , \ \ \ 
|\widetilde{d}_G| \simeq 1.4\times10^{-3} ~~~ (\lamNP \simeq 20~TeV) 
\ \ ,
\label{d-rate1}
\eq
for $100 \lsim \mh \lsim 200GeV$. The importance of this LHC
measurement, which in fact should precede the aforementioned one
at a Linear Collider, should be appreciated.  

The sensitivity to the 
NP couplings will  be notably improved if at LHC we also
study the shape of the $H+jet$ distributions at large transverse
momentum or large invariant masses. 
The change which is induced in the shape of these distributions 
constitutes a clear signal for NP. We have explored this possibility
by considering the ratio of the number of
events at large transverse momentum or invariant mass to the one at low
values. This quantity is rather insensitive to theoretical
uncertainties (due to  SM itself or to anomalous $gtt$ or $Htt$
couplings), but the precision is now controlled by the ~smallness
of the expected statistics. 
If only the $H\to\gamma\gamma$ mode is used, we reach about the same
sensitivity as the one given in (\ref{d-rate1}). 
This is already a very interesting result as
it constitutes a theoretically clean independent measurement. 
But now, depending of the value of the Higgs mass, the 
sensitivity  can be notably improved
by using the $WW^*$ and $ZZ^*$ modes, 
as well as the real $WW$ and $ZZ$ modes as soon as $\mh$  goes
above the corresponding thresholds. Depending on the value of
the Higgs mass, 
we then  expect an improvement by a factor 2 to 10
in these observability limits.\par

The values of the NP scales that such an analysis should allow to reach
\ie\@ 50 to 100 TeV,
are remarkable. They are of the same order of magnitude as the ones
expected in the electroweak sector for the operators 
${\O}_{WW}$, $\widetilde{\O}_{WW}$,
${\O}_{BB}$ and $\widetilde{\O}_{BB}$ describing anomalous $HZZ$,
$HZ\gamma$ and $H\gamma\gamma$ couplings and which affect the
corresponding $H$
decay modes as well as $HZ$,
$H\gamma$ production in $e^+e^-$ collisions and $H$ 
production in $\gamma\gamma$ collisions, 
\cite{Tsirigoti-CP, HZ,Vlachos, Helec}.\par 

A comparison of the effects or of the limits obtained in these two
(electroweak and gluonic) sectors should tell
us about the flavour and colour content of new physics.

\vspace*{0.5cm}
\noindent
\underline{Acknowledgments}\\
We like to thank Abdelhak Djouadi and Michael Spira  for
enjoyable and ~useful discussions, and G. Tsirigoti for 
participating in the early stage of the present work.

\newpage
\renewcommand{\theequation}{A.\arabic{equation}}
\setcounter{equation}{0}
\setcounter{section}{0}

{\large \bf Appendix A1: 1-loop SM contribution to the
$gg  H $ amplitude} \\

The form factor $\bar{f}(k^2)$ determining, (through 
the diagrams in Fig.1a), the Hgg coupling in SM 
in the case that the momentum $k$ of
one of the gluons may be off-shell, is

\bq
\bar{f}(k^2)~=~C_0(k^2) \left [1-\frac{4 \mt^2}{\mh^2-k^2}
\right ] -  \frac{2}{\mh^2-k^2} +  
\frac{2 k^2}{(\mh^2-k^2)^2} F(k^2)
\ ,
\eq
where all conventions are given in Sect.2.2. The triangle loop
is  computed through the 
FF-a package \cite{Oldenborgh}
using the standard Passarino-Veltman method \cite{PV}
and the notations of \cite{ABCDHag}
and \cite{Mertig}
with
\bq
C_0(k^2) \equiv C_0(k^2,0,\mh^2;\mt,\mt,\mt) \ .
\eq

and
\bq
     F(k^2)  \equiv  \int_0^1 dx\,
\ln \frac{\left[(1-x)m^2_1 +xm^2_2 -x(1-x)\mh^2 -i\epsilon
\right]}{\left[(1-x)m^2_1 +xm^2_2 -x(1-x)k^2 -i\epsilon
\right]} 
\eq

In the limit $k^2 \to 0$, in which both gluons are on-shell, 
$\bar{f}(0)$ matches the result
\cite{ellis, Hjet, RKEllis} 
\bq
\bar{f}(0)= -\, { \tau_t \over 2 m^2_t}(1+(1-\tau_t)\tilde f(\tau_t)) \ ,
\eq
where  $\tau_t= 4\mt^2/\mh^2$ and
\bqa
\tilde f(\tau_t) = \left [sin^{-1}(1/\sqrt{\tau_t})\right]^2 \ \ \ \ \ \ \ \
\ \ &
\makebox{\ \ \ \ \ if \ \ \ } &
           \tau_t \geq 1 \ \ \ \ , \ \ \ \nonumber \\[0.5cm]
\tilde f(\tau_t)= -{1\over4}\left [\ln \left ({1+\sqrt{1-\tau_t} 
\over 1-\sqrt{1-\tau_t}}
\right )
-i\pi\right ]^2 & \makebox{\ \ \ \ if \ \ \ } & \tau_t < 1 \ \
\ \ . \ \ \ \ 
\eqa

\vspace*{1.0cm}

{\large \bf Appendix A2 : CP conserving amplitudes for $gg\to Hg$}
\\

Below we discuss the 1-loop SM contribution, as well as the CP
conserving NP contribution, to the invariant amplitudes
for the process
\bq
g(\epsilon,k,a) + g(\epsilon',k',b) \to H(q) + g(e,q',c) \ ,
\label{ggHg}
\eq
where the polarization, momenta and colour indices of the gluons
are indicated in ~parentheses.
The  1-loop SM contribution to this amplitude
has been computed long ago by R.K. Ellis \etal\@ \cite{RKEllis}. 
It  arises from the triangle and
box terms appearing in Fig.1 and involved in Fig.3. 
We have recomputed their result in order to make sure that 
the NP contribution is added correctly to the SM one.\par

If Bose-symmetry among the three external gluons were ignored,
there would had been 14 different Lorentz invariant and CP
conserving forms contributing to the amplitude, 
which are  reduced to 
just  4 ones, when gauge symmetry is imposed. 
For the gluon momenta and
polarizations indicated in (\ref{ggHg}), these are 
\bqa
N_1& = & [(e.k)(k'.q')-(e.k')(k.q')] 
\{(k.q')[(\epsilon.\epsilon')(k'.q')
-(\epsilon.k')(\epsilon'.q')]\nonumber\\
&&+(\epsilon'.q')(\epsilon.q')(k.k')
-(\epsilon.q')(\epsilon'.k)(q'.k')\}\ ,
\eqa
\bq
N_2=[(\epsilon.e)(k.q')-(\epsilon.q')(e.k)]
[(\epsilon'.k)(k'.q')-(\epsilon'.q')(k.k')] \ , 
\eq
\bq
N_3=[(\epsilon'.e)(k'.q')-(\epsilon'.q')(e.k')]
[(\epsilon.k')(k.q')-(\epsilon.q')(k.k')] \ ,
\eq
\bqa
N_4 &=& [(e.k)(k'.q')-(e.k')(q'.k)]
[(\epsilon.k')(k.q')-(\epsilon.q')(k.k')] \cdot
\nonumber \\
&&[(\epsilon'.k)(k'.q')-(\epsilon'.q')(k.k')] \  .
\eqa
As a result the total contribution to the amplitude is written 
as\footnote{The $R$ amplitude has the phase of the $S$-matrix
elements. We use the same conventions for the couplings as in 
\cite{gunion, books}.} 
\bq
R(gg\to Hg)= - f_{abc}\sum_{j=1}^4 A_j(\hat{s},\hat{t},\hat{u}) N_j \ ,
\label{R(ggHg)}
\eq
where
\bqa
\hat s = (k+k')^2, & \hat t = (q-k)^2, &\hat u = (q'-k)^2 \ \ ,
\eqa
with the momenta defined in (\ref{ggHg}).
The $A_j$ amplitudes in (\ref{R(ggHg)}) are not the same as the 
ones used in \cite{RKEllis}. In accordance with this reference,
the requirements of
Bose-Einstein statistics among the three
external gluons reduce the number of these amplitudes to two, 
by determining  $A_2$, $A_3$ in terms of $A_1$, through
\bqa
A_2(\hat s, \hat t, \hat u ) &=& \frac{\hat t \hat s}{2 \hat u}
A_1(\hat u, \hat t, \hat s) \ \ ,  \label{ap-A2} \\
A_3(\hat s, \hat t, \hat u ) &=& -\, \frac{\hat u \hat s}{2 \hat t}
A_1(\hat t, \hat s, \hat u) \ \ ,  \label{ap-A3} 
\eqa
and imposing the constraints
\bqa
A_1(\hat s, \hat t, \hat u) & =& A_1(\hat s, \hat u, \hat t) \ \
, \label{SymA1} \\
A_4( \hat s , \hat t, \hat u) &=& A_4(\hat s, \hat u, \hat t)
\ \  , \label{SymA4} \\
A_4(\hat s, \hat t, \hat u)-A_4(\hat t, \hat s, \hat u)
& =& \frac{2}{\hat t} A_1(\hat t , \hat s, \hat u) -
    \frac{2}{\hat s} A_1(\hat s , \hat t, \hat u) \ \ .
\label{SymA1A4}
\eqa\par

Thus, we only need to give the 1-loop SM and the tree level
$\O_{GG}$ contributions to $A_1$ and $A_4$, which are
\bq
A_1(\hat s, \hat t, \hat u)=\frac{2 g^3_s m^2_t}{\pi^2 v \hat{t}\hat{u}^2}
\left [ 
{\hat{t}-\mh^2 \over \hat{t}}\bar f(\hat{t})-
{\hat{u}\over \hat{s}}\bar f(\hat{s})+{1\over2}\bar b_1
\right ]
+\left ({32g_s d_G\over v}\right ){(\hat{u}+\hat{s})(\hat{t}+
\hat{s})\over \hat{s}\hat{t}^2\hat{u}^2} \ ,
\label{ap-A1}
\eq
\bq
A_4(\hat s, \hat t, \hat u)=\frac{2 g^3_s m^2_t}{\pi^2 v \hat{t}\hat{u}^2}
\left [
{2\over \hat{t}}f(\hat{t})+{1\over2}\bar b_8
\right ]
- {64 g_s d_G\over v \hat t^2 \hat u^2} \ ,
\label{ap-A4}
\eq 
where  $\bar b_1$ and $\bar b_8$,
in (\ref{ap-A1}, \ref{ap-A4})
arise from the SM box diagram in Fig.1b. They are expressed in
terms of the $D$ functions defined in \cite{ABCDHag}
and  \cite{Mertig} as\footnote{See in particular Eqs.(D.12) of 
\cite{ABCDHag}.}

\bqa
D_i & = & D_i(0,0,0,\mh^2, \hat u, \hat s; \mt,\mt,\mt,\mt)\ , \\
\ol{D}_i & = & D_i(0,0,0,\mh^2, \hat u, \hat t; \mt,\mt,\mt,\mt) \ , \\
\ol{\ol{D}}_i & = & D_i(0,0,0,\mh^2, \hat t, \hat s; \mt,\mt,\mt,\mt) \ .
\eqa

\bqa
\bar b_1 & = & \mt^2 (D_0+2D_{13} -2\Dbb_{12} +\Db_0)
\nonumber \\
&& + \hat t (-2D_{133}
+2\Dbb_{122}-D_{25}-2D_{23}-\Db_{26}+2\Dbb_{22}+
2\Dbb_{24}-D_{13}+2\Dbb_{12}) 
\nonumber \\
&& + \hat u (-2D_{123}+2\Dbb_{123}-D_{24}-2D_{26}-2D_{25}-
\Db_{24}+2\Dbb_{26}
\nonumber \\
&& -\frac{1}{2}D_0 -D_{12}-2D_{13}-\frac{1}{2}\Db_0
-\Db_{12}- \frac{1}{2} \Dbb_{0}) 
\nonumber \\
&& + \hat s (-2D_{233}+2\Dbb_{223}-D_{26}-2D_{23}-\Db_{25}+
2\Dbb_{26}-D_{13}+ \Dbb_{12})
\nonumber \\
&& - 4D_{003}-8\Db_{003}+4\Dbb_{002}
-4D_{27}-4\Db_{27}+4\Dbb_{27} \ ,
\eqa
\bqa
\bar b_8 & = & 8 (D_{233}- \Db_{133}-\Dbb_{223} +D_{23}
-\Db_{23} -\Dbb_{26}) +2 (D_{13}-\Db_{13}-\Dbb_{12}) \ .
\eqa \par
\noindent
and computed using the FF-package \cite{Oldenborgh}.\\
We have checked (numerically) that all gauge non-invariant
terms arising from
the individual  diagrams in Fig.3, cancel out when they are 
added. A further test is provided
by the various Bose statistics relations (\ref{SymA1}-\ref{SymA1A4}) and
(\ref{ap-A2}, \ref{ap-A3}), which are also
perfectly satisfied. \par

\vspace*{1.0cm}

{\large {\bf Appendix A3 : Kinematics for H+jet production in
the parton model}}\\

The basic parton model expression for ~the hadron-hadron
collision $AB\to H f ...$, taking place through the subprocesses $a+b\to
H+f$ and illustrated in Fig.2,  is written as \cite{books}
\bq
\sigma(A B \to H ~f~ ...) = \sum_{ab} \int\!\int 
dx_adx_b~f_{a/A}(x_a)f_{b/B}(x_b) \hat{\sigma}(ab\to H f)
\ ,
\label{parton}
\eq
with $f$ being a gluon or massless quark jet.
Here $f_{a/A}(x_a)$ is the distribution function of partons of type
$(a=g,\, q,\, \bar q )$, in the hadron of type $A$.\par

We next list a few formulae for the kinematics.
The  transverse momenta of the produced $H$ and the $f$-jet
$p_T\equiv p_{TH}=p_{Tf}$ and the energy $E_{HT}=\sqrt{p^2_T+\mh^2}$
are described through
\bqa
x_T={2E_{HT}\over\sqrt{s}}, & \beta_T=p_T/E_{TH}
=\sqrt{1-{4\mh^2\over s
x^2_T}}, & x_{fT}={2E_{fT}\over\sqrt{s}}
=\beta_T x_T   \ .
\label{pT}
\eqa  
The rapidities of the Higgs and the outgoing jet $f$, 
in the laboratory system are related to their energies and
momenta along the beam-axis of hadron $A$, (taken as the 
$\hat z$-axis) and to the corresponding production angles
by 
\bqa
e^{2y_H} &= & {E_H+p_H\cos\theta_H\over
E_H-p_H\cos\theta_H} \ , \\
e^{2y_f} & = & {E_f+p_f\cos\theta_f\over 
E_f-p_f\cos\theta_f}  \ .
\eqa
The center-of-mass rapidity $\bar y$ of the  
$Hf$ pair, and the rapidities $y_H^*$, $y_f^*$ in
c.m., are defined as
\bq
y_H=\bar{y}+y^*_H, ~~~~~y_f=\bar{y}+y^*_f \  .
\label{cm-rapidity}
\eq
The fractions of momenta carried by 
the incoming partons are  given by 
\bqa
x_a & = & {x_T\over2}[ e^{y_H}+\beta_Te^{y_f}]=\frac{M}{\sqrt
s}\,  e^{\bar y}  \ , \\
x_b& =& {x_T\over2}[e^{-y_H}+ \beta_T e^{-y_f}]= \frac{M}{\sqrt
s} \, e^{-\bar y} \ , \eqa
while the Mandelstam invariants of the subprocesses  satisfy
\bq
\label{shat}
\hat{s}\equiv M^2=(p_a+p_b)^2= x_a x_b s=
E_{TH}^2 [1+\beta_T^2+2\beta_T \cosh(\Delta y)] \ ,
\eq
\bqa
\hat{t}=(p_H-p_a)^2 &=& -E_{TH}^2 \beta_T(\beta_T+e^{-\Delta y}
)\ , \label{that}\\
\hat{u}=(p_f-p_a)^2& =& -E_{TH}^2 \beta_T(\beta_T+e^{\Delta y}
) \ , \label{uhat}
\eqa
\bq
\tau~=~ \frac{\hat s}{s}~=~ x_ax_b  \  \ ,
\eq
where
\bq
\Delta y \equiv y_H-y_f = y_H^*-y_f^* \ \ .
\eq
Denoting the c.m. $Higgs$ particle production angle by $\theta^*$ and the
c.m. $H$ velocity by 
\bq
\beta=p_H^*/E_H^* =~ {\hat s-\mh^2 \over \hat s +m^2_H}\ ,
\label{beta}
\eq 
we have also the expressions 
\bqa
M &=& E_{TH} (1+\beta)\cosh y_H^*=  p_T (\frac{1}{\beta} +1)
\cosh y_f^* \ , \\
\hat t &=&  \frac{(\mh^2-\hat s)}{2}(1-\cos \theta^*) \ , \\
\hat u &=&  \frac{(\mh^2-\hat s)}{2}(1+\cos \theta^*) \ , \\ 
 \cos\theta^*& =& \frac{\tanh y_H^*}{\beta}=-\tanh y_f^* \ , \\
\chi\equiv e^{2y_H^*}&= & {\hat{u}-\mh^2 \over
\hat{t}-\mh^2}= {1+\beta \cos\theta^*\over 1- \beta \cos\theta^*}
\label{chi} \ .
\eqa\par

\vspace*{0.5cm}
{\bf The basic triple distribution} is obtained from
(\ref{parton}) to be
\bq
{d\sigma\over dp^2_T dy_H dy_f} =\tau S_{ij}
\eq
where, according to (\ref{parton}), $S_{ij}$ is the total 
probability
\bq
S_{ij}\equiv S_{gg} +S_{gq}+S_{g\bar q}+S_{q\bar q}
\eq
\noindent
with
\bq
S_{gg}\equiv g(x_a)g(x_b)\frac{d\hat{\sigma}(gg\to Hg)}{d \hat
t} \ ,
\label{Sgg}
\eq
\bqa
S_{gq} &\equiv & 
\sum_q \left [g(x_a)q(x_b){d\hat{\sigma}(gq\to Hq)\over d\hat{t}}
+q(x_a)g(x_b)
{d\hat{\sigma} (qg\to Hq) \over d\hat{t}} \right ] \ , 
\label{Sgq} \\
S_{g\bar q} & \equiv & \sum_{\bar q} 
\left [g(x_a)\bar q(x_b)
{d\hat{\sigma}(g\bar q\to H\bar q) \over d\hat{t}}
+\bar q(x_a)g(x_b)
{d\hat{\sigma} (\bar q g\to H\bar q) \over d\hat{t}}
\right ] \ , 
\label{Sgqbar} \\
S_{q\bar q} & \equiv & \sum_{q} 
\left [q(x_a)\bar q(x_b){d\hat{\sigma}(q\bar q\to Hg) \over d\hat{t}}
+\bar q(x_a)q(x_b)
{d\hat{\sigma} (\bar q q\to Hg)\over d\hat{t}} \right ] \ .
\label{Sqqbar}
\eqa
where $q=u,d,s,c,b$. From this basic distribution and 
imposing the cuts
\bq
|y_H| \leq Y_H, ~~~ |y_f| \leq Y_f \ , 
\eq
we get:\\  

\vspace*{0.5cm}
\noindent
{\bf The transverse energy distribution}\\
\bq
{d\sigma\over dx_T} =\int dy_H \int dy_f {M^2 x_T\over2}
S_{ij} \ ,
\eq
where $M^2$ is determined from (\ref{shat}) and the
integration limits are
\bqa
\label{yf-limits}
y_{fmin}&=& \max \Bigg \{ \ln\left ( {\beta_T x_T\over2-x_T e^{-y_H}}
\right ); -Y_f \Bigg \} \ , \nonumber \\
y_{fmax} & = & \min\Bigg \{\ln \left ({2-x_T e^{y_H}\over \beta_T x_T}
\right ); Y_f\Bigg \} \ , 
\eqa
\bqa
 && y_{Hmax}=-y_{Hmin}= \nonumber \\
&& \min\Bigg \{Y_H;~
\ln({2\over x_T});~ \cosh^{-1}\left (\frac{1}{x_T}(1+{m^2_H\over
s})\right ); \ln\left ({2-\beta_T x_T  e^{-Y_f}\over x_T}
\right ) \Bigg \} \ . 
\eqa\par

\vspace*{0.5cm}
\noindent
{\bf The rapidity distribution}\\
Since this is symmetric in $y_H$, we may consider
\bq
{d\sigma\over d|y_H|} =\int dx_T \int   dy_f M^2x_T
S_{ij} \ ,
\eq
where the $y_f$-limits are given (\ref{yf-limits}),
while the limits of the $x_T$ integration are
\bqa
x_{Tmin}& =& x_{\rm Tmin,exp} > {2m_H\over\sqrt{s}} \ ,
\nonumber \\ 
x_{Tmax}& = & \min \Big 
\{{1+{m^2_H\over s}\over \cosh y_H};~ {2e^{2Y_f\pm
y_H}-\sqrt{\Delta_1}\over e^{2(Y_f\pm y_H)}-1};~ 
2e^{\pm y_H}\Big \}
\eqa 
with
\bq
\Delta_1=4[e^{2Y_f}+{m^2_H\over s}(1-e^{2(Y_f-y_H)})]
\ .
\eq\par

\vspace*{0.5cm}
\noindent
{\bf The invariant mass distribution}\\
Using the c.m. rapidity $\bar{y}$ defined above and (\ref{chi}),
one obtains
\bq
{d\sigma\over dM^2} =\int d\chi\int  d\bar{y}\,
{(M^2+m^2_H) \over s(1+\chi)^2} S_{ij} \ ,
\eq
where the integration limits are
\bqa
\bar{y}_{max} & = & \min \Big 
\{Y_H-{1\over2}\ln\chi;~ Y_f-{1\over2}\ln \left({M^2-\chi
m^2_H\over M^2\chi-m^2_H}\right );~ \ln({\sqrt{s}\over M}) 
\Big \} \ , \nonumber \\
\bar{y}_{min} & = & \max \Big \{-Y_H-{1\over2}\ln\chi;~
 -Y_f-{1\over2}\ln \left ({M^2-\chi
m^2_H\over M^2\chi-m^2_H}\right );~ 
 -\ln({\sqrt{s}\over M}) \Big \} \ , \label{ybar} \\
\chi_{max}& =& \min\Big 
\{{M^2\over m^2_H}; 
\frac{M^2(s+m^2_H e^{-2Y_f})}{M^4e^{-2Y_f}+sm^2_H};
\frac{m^2_H(1-e^{2(Y_H+Y_f)})+\sqrt{\Delta_2}}{2M^2};
\frac{s}{M^2}e^{2Y_H}\} , \nonumber \\
\chi_{min}& =& \max\Big 
\{{m^2_H\over M^2}; 
\frac{M^4e^{-2Y_f}+sm^2_H}{M^2(s+m^2_H e^{-2Y_f})}; 
\frac{2M^2}{m^2_H(1-e^{2(Y_H+Y_f)})+\sqrt{\Delta_2}};
\frac{M^2}{s}e^{-2Y_H} \Big \} ,
\eqa
where
\bq
\Delta_2 = \mh^4\left ( e^{2(Y_H+Y_f)}-1 \right )^2+
4 M^4e^{2(Y_H+Y_f)} \ .
\eq

\vspace*{0.5cm}
\noindent
{\bf The angular distribution} is given by
\bq
{d\sigma\over d\chi} =\int dM^2 \int   d\bar{y} 
\, {(M^2+m^2_H) \over s(1+\chi)^2} S_{ij} \ .
\eq
The $\bar y$ integration limits are as in (\ref{ybar}), while
for the $M^2$ integration we have the limits
\bqa
M^2_{max}& =& \min
\Big \{\chi s e^{2Y_H}; {s\over\chi}e^{2Y_H}; M^2_+;
M^{'2}_+; s \Big \} \ , \nonumber \\
M^2_{min}& =& \max
\Big \{{\chi m^2_H(e^{2(Y_f+Y_H)}-1)
\over\chi^2e^{2(Y_f+Y_H)}-1}; {\chi
m^2_H(e^{2(Y_f+Y_H)}-1)\over e^{2(Y_f+Y_H)}-\chi^2}; M^2_-;
M^{'2}_-\Big \} \ , 
\eqa
where
\bqa
M^2_{\pm}& =&
{1\over2}[\chi(m^2_H+se^{2Y_f})\pm\sqrt{\Delta_3}], \\
\Delta_3& =& \chi^2(m^2_H+se^{2Y_f})^2-4sm^2_He^{2Y_f}, \\
M'^2_{\pm}&= &{1\over2\chi}[m^2_H+se^{2Y_f}\pm\sqrt{\Delta_3'}]
, \\
\Delta_3'&= & (m^2_H+se^{2Y_f})^2-4sm^2_H\chi^2e^{2Y_f} .
\eqa

\newpage

\begin{center}

{\large \bf Figure captions}
\end{center}
\vspace{0.5cm}

\noindent
{\bf Fig.1} Diagrams for top quark loop SM contributions to $Hgg$ and
$Hggg$ couplings. \\

\noindent
{\bf Fig.2} Diagrams for $p+p\to H+jet+X$. \\

\noindent
{\bf Fig.3} Diagrams for $gg\to Hg$, $gq\to Hq$ and $q\bar q\to Hg$.\\

\noindent
{\bf Fig.4} Rapidity distribution in $p+p\to H+jet+X$ at LHC, for
$\mh=100~GeV$.
$SM$ describes 1-loop SM predictions; $SM_{eff}$
the large $m_t$ approximation to SM; $+d_G$ and $-d_G$
describe the ${\O}_{GG} $ contributions for 
$d_G=+ 10^{-3}$ and $d_G=- 10^{-3}$
respectively; and $\wtil{d}_G$ describes the $\widetilde{\O}_{GG}$
contribution for $|\widetilde{d}_G|=10^{-3}$.\\

\noindent
{\bf Fig.5} Angular distribution in $p+p\to H+jet+X$ at LHC.
See caption of Fig.4.\\

\noindent
{\bf Fig.6} Transverse energy distribution in $p+p\to H+jet+X$
at LHC. See caption of Fig.4.\\

\noindent
{\bf Fig.7} Invariant mass distribution in $p+p\to H+jet+X$ at
LHC. See caption of Fig.4.\\


\begin{figure}[p]
\vspace*{-4cm}
\hspace*{-2.cm}\epsfig{file=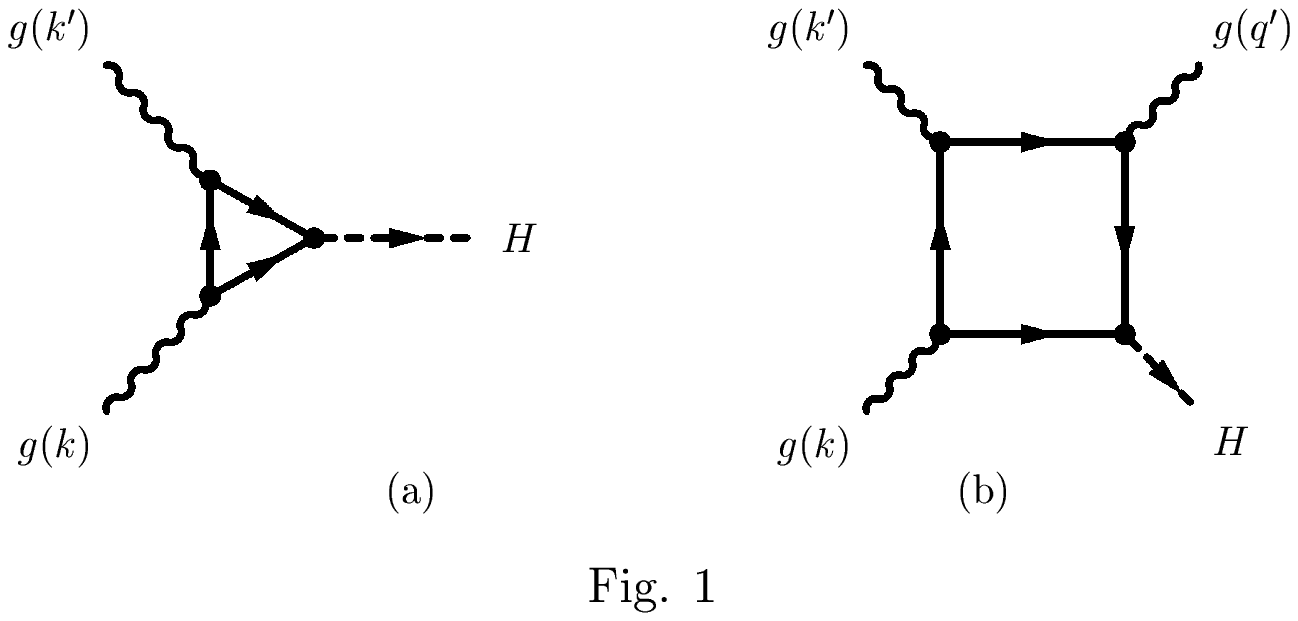}
\vspace*{-19cm}
\begin{center}
Diagrams for top quark loop SM contributions to $Hgg$ and
$Hggg$ couplings. 
\label{fig1}
\end{center}
\end{figure}
\vspace*{1cm}
\begin{figure}[p]
\vspace*{-4cm}
\hspace*{-2.cm}\epsfig{file=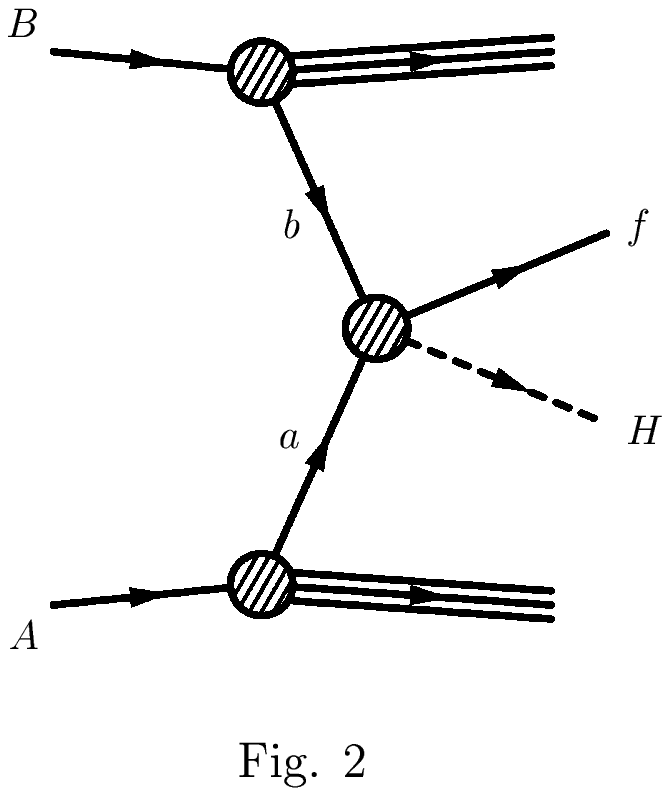}
\vspace*{-17.cm}
\begin{center}
Diagrams for $p+p\to H+jet+X$. 
\label{fig2}
\end{center}
\end{figure}
\newpage
\begin{figure}[h]
\vspace*{-4cm}
\hspace*{-2.5cm}$$\epsfig{file=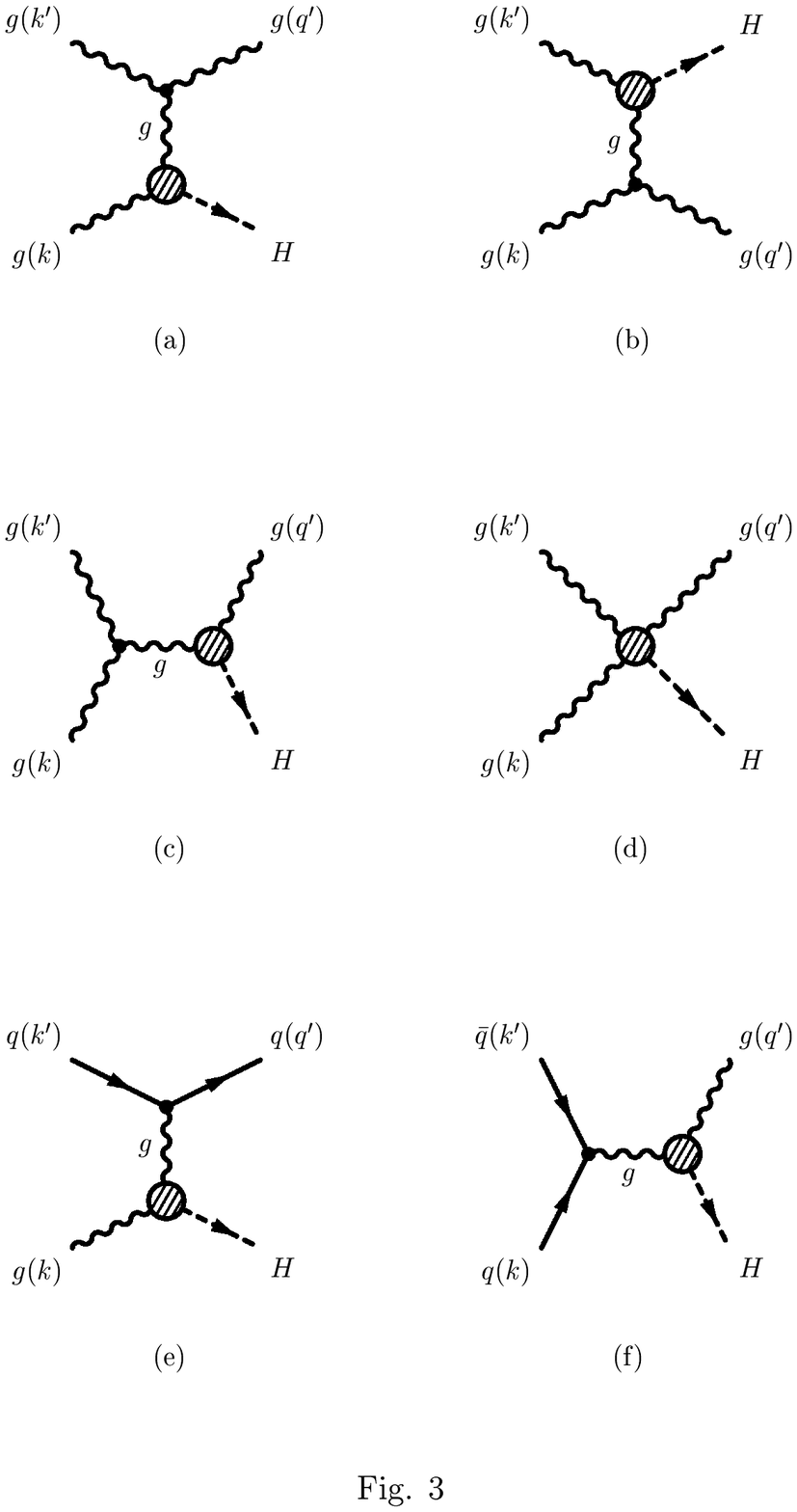,height=22cm}$$
\vspace*{-3.cm}
\begin{center}
Diagrams for $gg\to Hg$, $gq\to Hq$ and $q\bar q\to Hg$. 
\label{fig3}
\end{center}
\end{figure}

\clearpage
\newpage

\begin{figure}[p]
\vspace*{-11cm}
\begin{center}
\hspace*{-4cm}
\mbox{
\epsfig{file=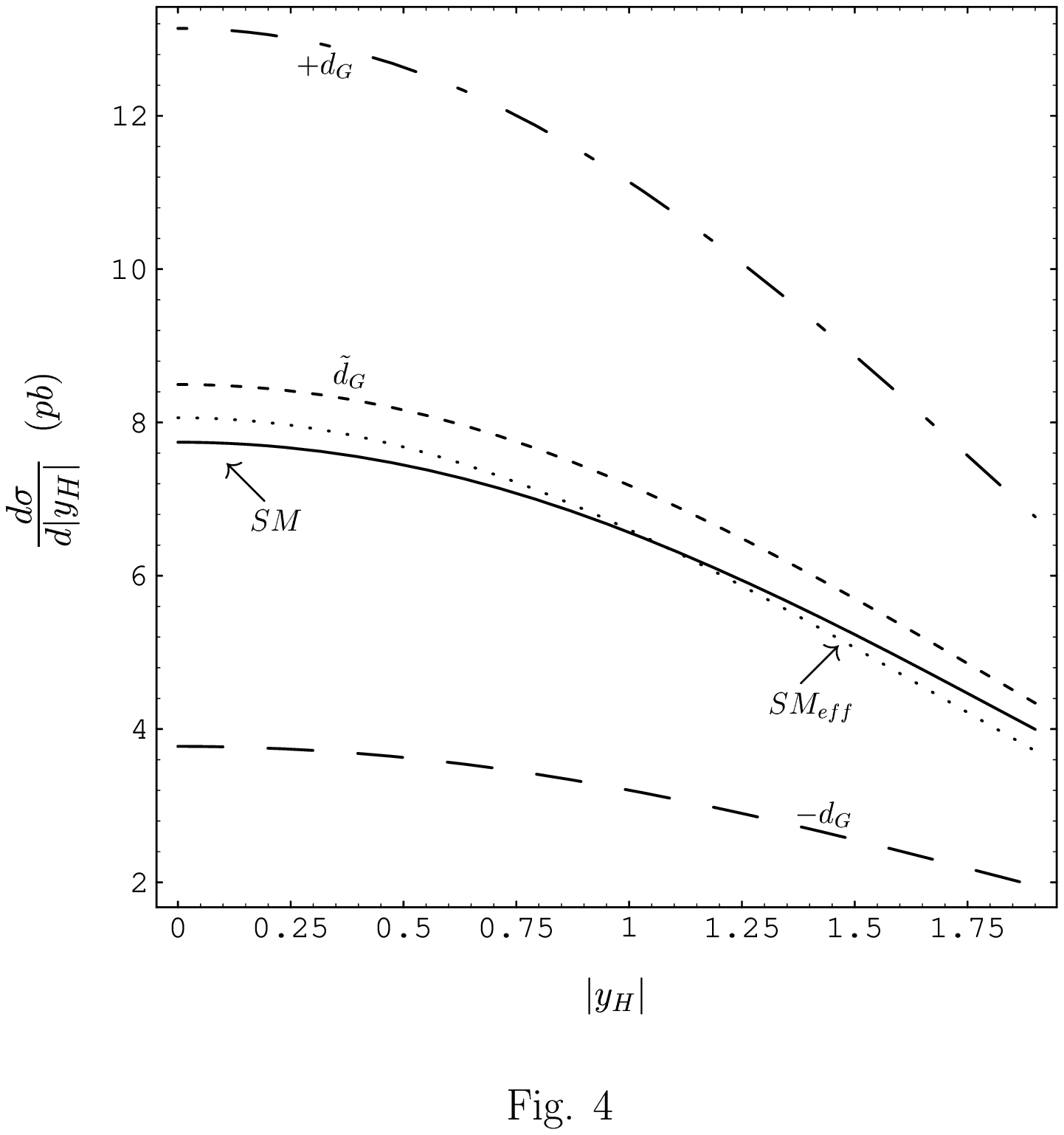,height=32cm}}
\vspace*{-8.cm}

\noindent
Rapidity distribution in $p+p\to H+jet+X$ at LHC, for
$\mh=100~GeV$. 
$SM$ describes 1-loop SM predictions; $SM_{eff}$
the large $m_t$ approximation to SM; $+d_G$ and $-d_G$
describe the ${\O}_{GG} $ contributions for 
$d_G=+ 10^{-3}$ and $d_G=- 10^{-3}$
respectively; and $\wtil{d}_G$ describes the $\widetilde{\O}_{GG}$
contribution for $|\widetilde{d}_G|=10^{-3}$.
\label{fig4}
\end{center}
\end{figure}

\clearpage
\newpage

\begin{figure}[p]
\vspace*{-11cm}
\begin{center}
\hspace*{-4cm}
\mbox{
\epsfig{file=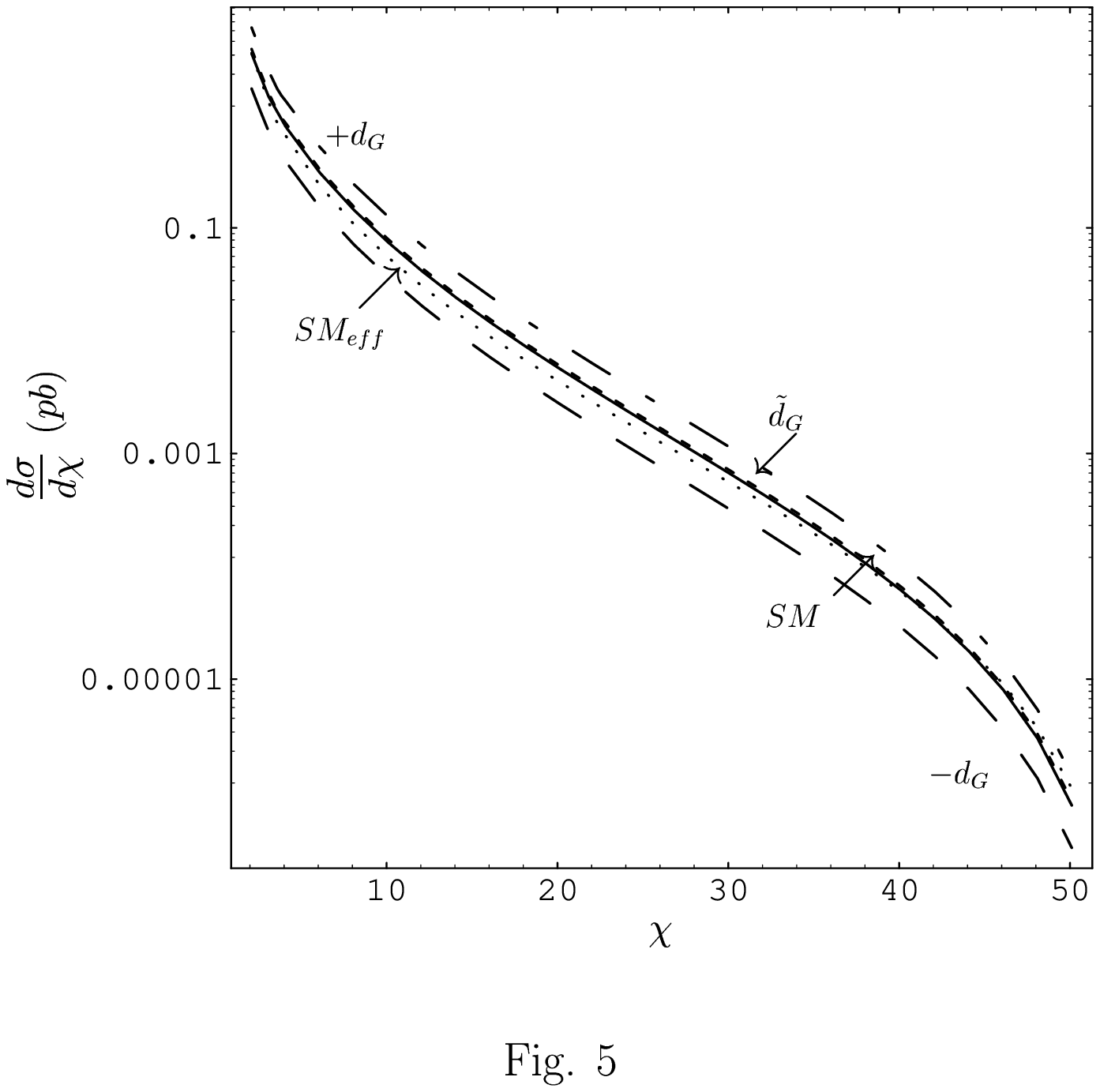,height=32cm}}
\vspace*{-9.cm}

Angular distribution in $p+p\to H+jet+X$ at LHC.
See caption of Fig.4. 
\label{fig5}
\end{center}
\end{figure}

\clearpage
\newpage

\begin{figure}[p]
\vspace*{-11cm}
\begin{center}
\hspace*{-4cm}
\mbox{
\epsfig{file=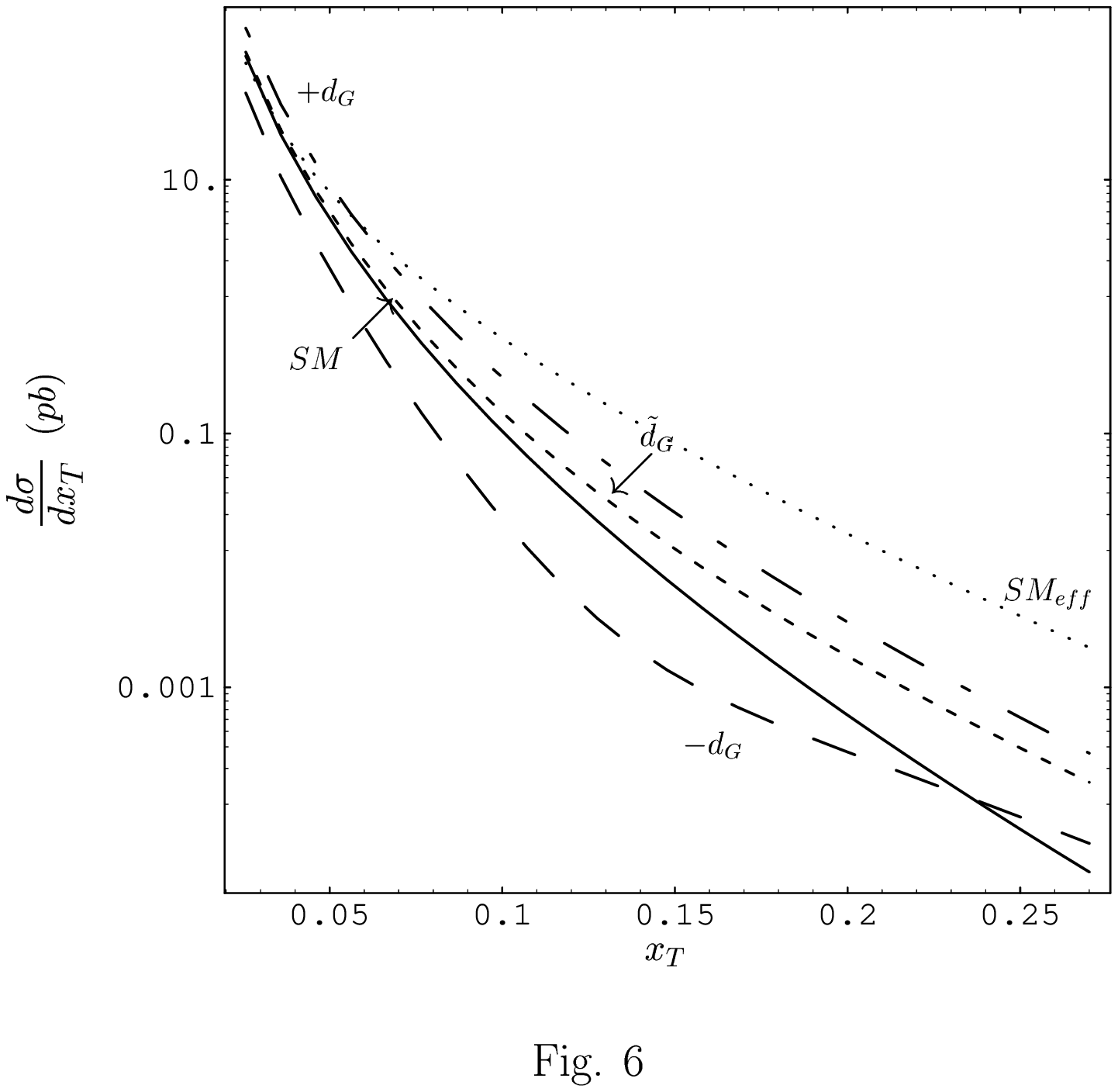,height=32cm}}
\vspace*{-9.cm}

Transverse energy distribution in $p+p\to H+jet+X$ at LHC. 
See caption of Fig.4.
\label{fig6}
\end{center}
\end{figure}

\clearpage
\newpage

\begin{figure}[p]
\vspace*{-11cm}
\begin{center}
\hspace*{-4cm}
\mbox{
\epsfig{file=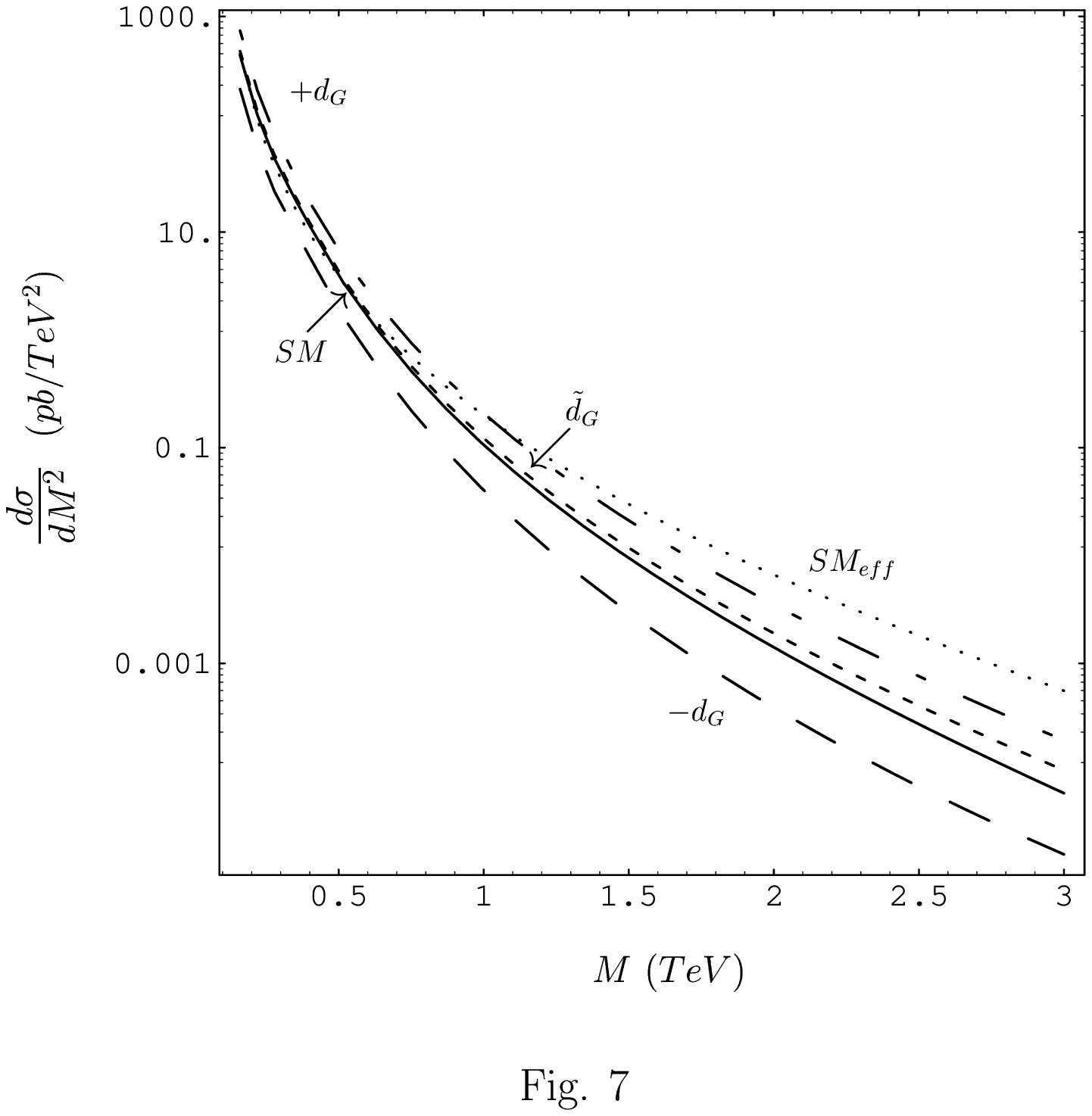,height=32cm}}
\vspace*{-8.cm}

Invariant mass distribution in $p+p\to H+jet+X$ at
LHC. See caption of Fig.4.
\label{fig7}
\end{center}
\end{figure}

\end{document}